\documentstyle[prcrc,11pt,fleqn,epsf]{article}

\newcommand{\mbh}{M_{\rm bh}}
\newcommand{\rh}{R_{\rm h}}
\newcommand{\dc}{\delta_{\rm c}}
\newcommand{\mh}{M_{\rm h}}
\newcommand{\ea}{et al.~}

\title{Numerical investigation of the threshold for primordial black
hole formation} 
\author{J.C. Niemeyer
\address{University of Chicago, Department of Astronomy and
Astrophysics,5640 S. Ellis Avenue, Chicago, IL 60637, USA}}

\begin{document}
\maketitle

\begin{abstract}
First results of a numerical investigation of primordial black hole
formation in the radiation
dominated phase of the Early Universe are presented. The simulations
follow the gravitational collapse of three
different families of high-amplitude density fluctuations imposed at the time
of horizon crossing. The threshold for black hole 
formation, $\delta_{\rm c} \approx 0.7$, is found to be nearly
identical for all perturbation families if the control parameter, $\delta$,
is chosen as the total excess mass within the initial horizon
volume. Furthermore, we demonstrate that the scaling of black hole mass
with distance from the formation threshold, known to occur in
near-critical gravitational collapse, applies to
primordial black hole formation. 
\end{abstract}

\section{Introduction}
Every quantitative analysis of the primordial black hole (PBH) number
and mass spectrum \cite{carr75}
requires knowledge of the threshold parameter, $\dc$, separating
perturbations that form 
black holes from those that do not, and the resulting black hole mass,
$\mbh$, as a function of distance from the threshold. In order to
determine $\dc$ and $\mbh$ for various initial conditions, 
we performed one-dimensional, 
general relativistic simulations of the hydrodynamics of PBH
formation in the radiation-dominated phase of the early
universe. Three families of perturbation shapes were
chosen to represent ``generic'' classes of initial data, reflecting the lack of
information about the specific shape of primordial fluctuations. The
numerical technique is sketched 
in Section \ref{numerics}. Defined as the excess mass within the
horizon sphere at the onset of the collapse, we find $\dc \approx 0.7$
for all three perturbation shapes, indicating that the threshold value
may indeed be universal (Section \ref{hydro}). A numerical confirmation of the
previously suggested power-law scaling of $\mbh$ with
$\delta - \dc$ \cite{niejed97}, related to the well-known behavior of
collapsing space--times at the critical point of black hole formation
\cite{chop93}, is presented in Section \ref{scaling}. In this framework,
the PBH mass spectrum is determined by the dimensionless coefficient,
$K$, the scaling exponent, $\gamma$, and the initial horizon mass,
$\mh$, such that
\begin{equation}
\label{scale}
\mbh = K \mh (\delta - \dc)^\gamma\,\,.
\end{equation}
We provide numerical results for $K$ and $\gamma$ for
the three perturbation families. A more detailed description of the
numerical technique and further results will be presented in a
forthcoming publication \cite{niejed98}. 

\section{Numerical technique}
\label{numerics}
The dynamics of collapsing density perturbations in the Early Universe
are determined by the general relativistic
equations of motion for a perfect fluid, the field equations, the first
law of thermodynamics, and a radiation-dominated equation of state. The
assumption of spherical symmetry is well justified for large
fluctuations in a Gaussian distribution \cite{barea86}, reducing the
problem to one spatial dimension. 

For our simulations, we chose the formulation of the hydrodynamical
equations by Hernandez and Misner \cite{hermis66} as implemented by
Baumgarte \ea \cite{baum95}. Based on the original equations by
Misner and Sharp  
\cite{missha64}, Hernandez and Misner proposed to exchange the
Schwarzschild time variable, $t$, with the outgoing null coordinate,
$u$. In so doing, the hydrodynamical equations retain the Lagrangian
character of the Misner--Sharp equations but avoid crossing into the
event horizon of a black hole once it has formed. Covering the entire
space--time outside while asymptotically approaching the event horizon,
the Hernandez--Misner equations are perfectly suited to
follow the evolution of a black hole for long times after its
formation without encountering coordinate singularities. This allowed
us, in principle, to study the
accretion of material onto newly formed PBHs for arbitrarily long times (in
contrast with earlier calculations \cite{Nade}).  
Since the expanding outer regions of our simulated piece of the
universe are most conveniently tracked in a comoving numerical
reference frame, the Lagrangian form of the Hernandez--Misner equations
is their second major asset. It also provides a simple prescription
for the outer boundary condition, which is defined to match the exact
solution of the Friedmann equations for a radiation 
dominated flat universe. Hence, the pressure follows the
analytic solution 
\begin{equation}
P = P_0 \left(\frac{\tau}{\tau_0}\right)^{-2}\,\,,
\end{equation}
where $\tau$ is the proper time of the outermost fluid element
(corresponding to the cosmological time $t$ in a
Friedmann--Robertson--Walker (FRW) universe) and
$P_0$ and $\tau_0$ are the initial values for pressure and proper
time.

\section{Threshold for black hole formation}
\label{hydro}
We studied the spherically symmetric evolution of three families of curvature
perturbations. Initial conditions were chosen to be
perturbations of the energy density, $\epsilon = \rho_0 e$, in
unperturbed Hubble flow specified
at horizon crossing. The first family of perturbations is described by
a Gaussian-shaped overdensity that asymptotically approaches the FRW
solution at large radii. The other two families of initial conditions
involve a spherical
Mexican Hat function and a fourth order polynomial. These functions
are characterized by
rarefaction regions outside of the horizon radius, $\rh$, that
identically compensate 
for the additional mass of the overdensities inside the horizon
volume, so that the mass derived from the total integrated 
density profile is equal to that of an unperturbed FRW solution.
In our numerical experiments, the amplitude, $A$, of the perturbations
is used to tune the 
initial conditions to sub- or supercriticality with respect to black
hole formation. The critical amplitude, $A_{\rm c}$, is
strongly shape-dependent, varying between $A_{\rm c} = 3.04$ for
Mexican-Hat-shaped perturbations and $A_{\rm c} = 2.05$ for the
Gaussian curve. If, however, we define the control
parameter $\delta$ as the additional mass 
inside $\rh$ in units of the horizon mass, we find strikingly similar
values --- $\dc = 0.67$ (Mexican Hat), $\dc = 0.70$ (Gaussian curve), and 
$\dc = 0.71$ (polynomial) --- for all three families of initial data in our
study. This number is considerably greater than the previously
employed threshold $\dc = 1/3$
following from analytical estimates \cite{carr75}. Given
the qualitative difference of the functional forms of the different
perturbation families, the result that $\dc$ of the Gaussian
perturbation lies in between
the critical values of the mass compensated functions is
surprising. Our results suggest that $\dc \approx 0.7$ (with $\delta$
defined as above) is a universal statement, i.e., true for all
perturbation shapes in the radiation-dominated regime. This remains to
be verified by means of additional experiments. 

\section{Scaling of PBH masses with distance to the threshold} 
\label{scaling}

For a variety of matter models, it is well-known that the dynamics of
near-critical collapse
exhibit continuous or discrete self-similarity and power-law scaling
of the black hole mass with the offset from the critical point
\{Eq.~(\ref{scale}) \cite{chop93,critrev}\}. In particular, Evans
and Coleman \cite{evans94} found 
self-similarity and mass scaling in numerical experiments of a
collapsing radiation fluid. They numerically determined the scaling exponent
$\gamma \approx 0.36$, followed by a linear perturbation analysis of
the critical solution by Koike \ea \cite{koike95} that yielded $\gamma \approx
0.3558$. Until recently, it was believed that entering the scaling regime
requires a degree of fine-tuning of the initial data
that is unnatural for any astrophysical application. It was noted
\cite{niejed97} that fine-tuning to criticality
occurs naturally in the case of PBHs forming from a steeply declining
distribution of primordial density fluctuations, as generically predicted by
inflationary scenarios. In the radiation-dominated
cosmological epoch, the only difference with the fluid collapse
studied by Evans and Coleman \cite{evans94} is the expanding,
finite-density-background space--time of a FRW universe. Assuming that 
self-similarity and mass scaling are consequences of an intermediate
asymptotic solution that is independent of the asymptotic boundary
conditions, Eq.~(\ref{scale}) is applicable to PBH masses,
allowing the derivation of a universal PBH initial mass function
\cite{niejed97}. 
\begin{figure}
\epsfysize=8cm
\epsfbox{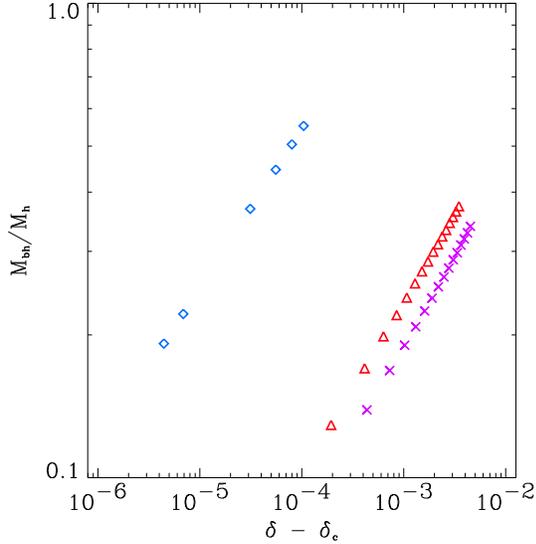}
\vspace{-13mm}
\caption{\label{f7} Black hole masses as a function of $\delta - \dc$
for three different 
perturbation shape families. The best fit parameters to equation
(\ref{scale}) are: 
$\gamma = 0.36$, $K = 2.85$, $\dc = 0.6745$ (Mexican Hat
perturbation, triangles); $\gamma = 0.37$, $K = 2.39$, $\dc = 0.7122$
(polynomial perturbation, crosses); $\gamma = 0.34$, $K = 11.9$,
$\dc = 0.7015$ (Gaussian curve perturbation, diamonds).}
\end{figure}
Figure (\ref{f7}) presents numerical evidence for mass scaling
according to Eq.~(\ref{scale}) in black hole collapse in an
asymptotic FRW space--time. All three perturbation families give rise
to scaling solutions with a scaling exponent $\gamma
\approx 0.36$.

\section{Conclusions}
This work discusses numerical collapse simulations of three
generic families of energy density perturbations, one with a finite
total excess mass
with respect to the unperturbed FRW solution and two mass compensated
ones. Among various possible definitions for the collapse control parameter
$\delta$, the total excess
gravitational mass of the perturbed space--time with respect to the
unperturbed FRW background enclosed in the initial horizon volume is
the only one that gives rise to a 
similar threshold value for all three shape families, $\dc \approx
0.7$. Whether this result is an indication for universality of $\dc$
in this specific definition needs to be verified with the help of
additional simulations using a larger sample of initial perturbation
shapes.  

The previously suggested \cite{niejed97} scaling relation between
$\mbh$ and $\delta - \dc$, based on the analogy with critical
phenomena observed in near-critical black hole collapse in
asymptotically non-expanding space--times \cite{chop93}, is confirmed
numerically for an asymptotic FRW background. For the smallest black
holes in this investigation, the scaling exponent is $\gamma \approx
0.36$, which is identical to the non-expanding numerical and
analytical results \cite{evans94,koike95} within our numerical accuracy. The
parameter $K$ of Eq.~(\ref{scale}), needed in addition to
$\gamma$ to evaluate the two-parameter PBH
IMF derived in \cite{niejed97}, ranges from $K \approx 2.4$ to $K
\approx 12$.  

The author wishes to thank K. Jedamzik for a fruitful collaboration,
T. Baumgarte for providing the original version of 
the hydrodynamical code, and Joan George for valuable stylistic corrections.


\begin{thebibliography}{99}
\bibitem{carr75}  B. J. Carr, ApJ 201, (1975) 1.

\bibitem{niejed97} J. C. Niemeyer and K. Jedamzik, Phys. Rev. Lett.,
in press (1998); see also astro-ph/9709072.

\bibitem{chop93} M. W. Choptuik, Phys. Rev. Lett. 70 (1993) 9.

\bibitem{niejed98} J. C. Niemeyer and K. Jedamzik, Numerical
Investigation of Primordial Black Hole Formation, in preparation.

\bibitem{barea86} J. M. Bardeen, J. R. Bond, N. Kaiser, and
A. S. Szalay, ApJ 304 (1986) 15.

\bibitem{hermis66} W. C. Hernandez and C. W. Misner, ApJ
143 (1966) 452.

\bibitem{baum95} T. W. Baumgarte, S. L. Shapiro, and S. A. Teukolsky,
ApJ 443 (1995) 717. 

\bibitem{missha64} C. W. Misner and D. H. Sharp, Phys. Rev. 136 (1964)
B571. 

\bibitem{Nade} D. K. Nadezhin, I. D. Novikov, and A. G. Polnarev,
Sov. Astron. 22 (1978) 129; see also G.~V.~Bicknell and R. N. Henriksen,
ApJ 232 (1979) 670.

\bibitem{critrev} C. Gundlach, gr-qc/9712084 (1997); see also M. W. Choptuik,
gr-qc/9803075 (1998).

\bibitem{evans94} C. R. Evans and J. S. Coleman, Phys. Rev. Lett.
72 (1994) 1782. 

\bibitem{koike95} T. Koike, T. Hara, and S. Adachi,
Phys. Rev. Lett. 74 (1995) 5170.

\end{thebibliography}
\end{document}